\documentclass[usenatbib]{mn2e}
\usepackage{epsfig}
\usepackage{graphicx}

\usepackage{color}
\usepackage{journals}

\newcommand {\bc}{\begin {center}}
\newcommand {\ec}{\end {center}}
\newcommand {\be}{\begin {equation}}
\newcommand {\ee}{\end {equation}}
\newcommand {\beq}{\begin {eqnarray}}
\newcommand {\eeq}{\end {eqnarray}}

\newcommand {\ergs}{{\rm erg\ \rm s^{-1}}}
\renewcommand{\d}{{\rm d}}
\newcommand{\mass}{m}
\def\ff {{\rm f}}
\def\ii {{\rm i}}
 
\title[The critical luminosity for magnetized neutron stars]
{The critical accretion luminosity for magnetized neutron stars}
\author[A. A.~Mushtukov et al.] {Alexander~A.~Mushtukov,$^{1,2}$\thanks{E-mail: al.mushtukov@gmail.com}  
Valery~F.~Suleimanov,$^{3,4}$
Sergey~S.~Tsygankov$^{1,5}$ 
\newauthor
and  Juri Poutanen$^{1}$ \\
$^1$Tuorla Observatory, Department of Physics and Astronomy, University of Turku, 
  V\"ais\"al\"antie 20, FI-21500 Piikki\"o, Finland \\
$^2$Pulkovo Observatory of Russian Academy of Sciences, Saint-Petersburg 196140, Russia \\
$^3$Institut f\"ur Astronomie und Astrophysik, Kepler Center for Astro and Particle Physics, Universit\"at T\"ubingen, Sand 1, D-72076 T\"ubingen, Germany \\
$^4$Kazan (Volga region) Federal University, Kremlevskaja str., 18, Kazan 420008, Russia \\
$^5$Finnish Centre for Astronomy with ESO (FINCA), University of Turku,  V\"ais\"al\"antie 20, FI-21500 Piikki\"o, Finland
}

\begin{document}

\pagerange{\pageref{firstpage}--\pageref{lastpage}}
\pubyear{2014}

\maketitle
\label{firstpage}

\begin{abstract}
The accretion flow around X-ray pulsars with a strong magnetic field is funnelled by 
the field to relatively small regions close to the magnetic poles of the neutron star (NS), the hotspots. 
During strong outbursts regularly observed from some X-ray pulsars,
the X-ray luminosity can be so high, that the emerging radiation is able
to stop the accreting matter above the surface via radiation-dominated shock, 
and the accretion column begins to rise. 
This border luminosity is usually called the ``critical luminosity''.
Here we calculate the critical luminosity as a function of the NS magnetic
field strength $B$ using exact Compton scattering cross section in strong magnetic field. 
Influence of the resonant scattering and photon polarization is taken into account for the first time. 
We show that the critical luminosity is not a monotonic function of the $B$-field. 
It reaches a minimum of a few $10^{36}$~erg~s$^{-1}$ when the cyclotron energy is about 10 keV
and a considerable amount of photons from a hotspot have energy close to the cyclotron resonance. 
For small $B$, this luminosity is about  $10^{37}$~erg~s$^{-1}$, nearly independent of the parameters. 
It grows for the $B$-field in excess of $10^{12}$~G because of the drop in the effective cross-section 
of interaction below the cyclotron energy. 
We  investigate how different types of the accretion flow and geometries of the accretion channel  affect the results and 
demonstrate that the general behaviour of the critical luminosity on $B$-field is very robust.
The obtained results are shown to be in a good
agreement with the available observational data and provide a necessary
ground for the interpretation of upcoming high quality data from the
currently operating and planned X-ray telescopes.
\end{abstract}


\section{Introduction}
\label{intro}

The strong magnetic field ($B$-field) with the strength as high as
$10^{12}$--$10^{13}$~G in X-ray pulsars (XRPs) strongly affects the accretion
process to the neutron star (NS). Namely, at some distance from the
NS, called magnetospheric radius, the magnetic pressure balances the
ram pressure of the infalling gas. At this point, plasma cannot move
across the magnetic field lines any more, and hence funnelled to the
relatively small regions on the NS surface close to the magnetic
poles, the hotspots, where releases its kinetic energy in X-rays.
Compactness of the hotspots (whose area could as small as
$10^{-5}$--$10^{-4}$ of the total NS surface) in
combination with the high mass accretion rate (occurring for example during giant
type II outbursts observed from XRPs with Be companions) 
lead to a strong radiation pressure force that is able
to stop the infalling matter above the NS surface. 
This happens at the so called {\it critical}  luminosity  \citep{BS1976}. With the
further increase of the mass accretion rate, and hence the luminosity, the
accretion column starts to rise above the hotspot. Therefore,
the critical luminosity divides two regimes of accretion onto a NS
with strong magnetic field. Below it, plasma reaches the NS surface heating it up via Coulomb collisions
\citep{ZSh1969}. At higher
luminosities, when the accretion column is expected to rise, the
accreted matter is decelerated in the radiation-dominated shock on top
of the column \citep{BS1976}.

Observational manifestation of dependence of the  accretion column height 
on the XRP luminosity is an anti-correlation of the cyclotron
absorption line energy with the observed source flux
(\citealt{TsL2006}; \citealt*{TsLS2010}). Such cyclotron absorption features (sometimes with
higher harmonics) observed in the energy spectra of XRPs \citep{Cob2002,Fil2005,CW2012} 
provide a standard method to estimate 
the magnetic field strength \citep{Gnedin1974}. 
Qualitative explanation of a negative correlation of the
cyclotron energy with luminosity in bright pulsars has been proposed by
different authors
(\citealt*{Mih2004}; \citealt{Poutanen2013,Nish2014}). Interestingly, in
the low-luminosity XRPs, a positive correlation of the cyclotron
line energy with flux was observed \citep{Staub2007,Yam2011,Kloch2012}. 
The models explaining this behaviour assume that in this case the
pulsar luminosity is below the critical one 
(\citealt{Staub2007}; \citealt*{Mukh2013}). 

Different behaviour of the ``cyclotron energy --
luminosity'' dependence gives a possibility to estimate the value of the
luminosity from  observations. The luminosity where
the positive correlation is changed by the negative one 
can be associated with the critical luminosity, where 
the radiation pressure is strong enough to stop the infalling matter. 
Measuring the value of the critical luminosity is extremely
important because it contains valuable information about interaction of
radiation with matter in strong $B$-field.

The value of the critical luminosity  is defined by processes which provide
radiation pressure. In a case of strongly magnetized NSs, 
it is mainly Compton scattering (see Section~\ref{sec:BasicIdeas}). 
Because the scattering cross-section in strong magnetic field has a rather 
complicated behaviour (it depends strongly on photon energy,
polarization state and the $B$-field strength, and includes a number of
resonances, see \citealt*{Her1982,Harding1986,Harding1991}), calculation
of the effective cross-section becomes a key problem.

\begin{figure*}
\centering 
\includegraphics[width=15cm]{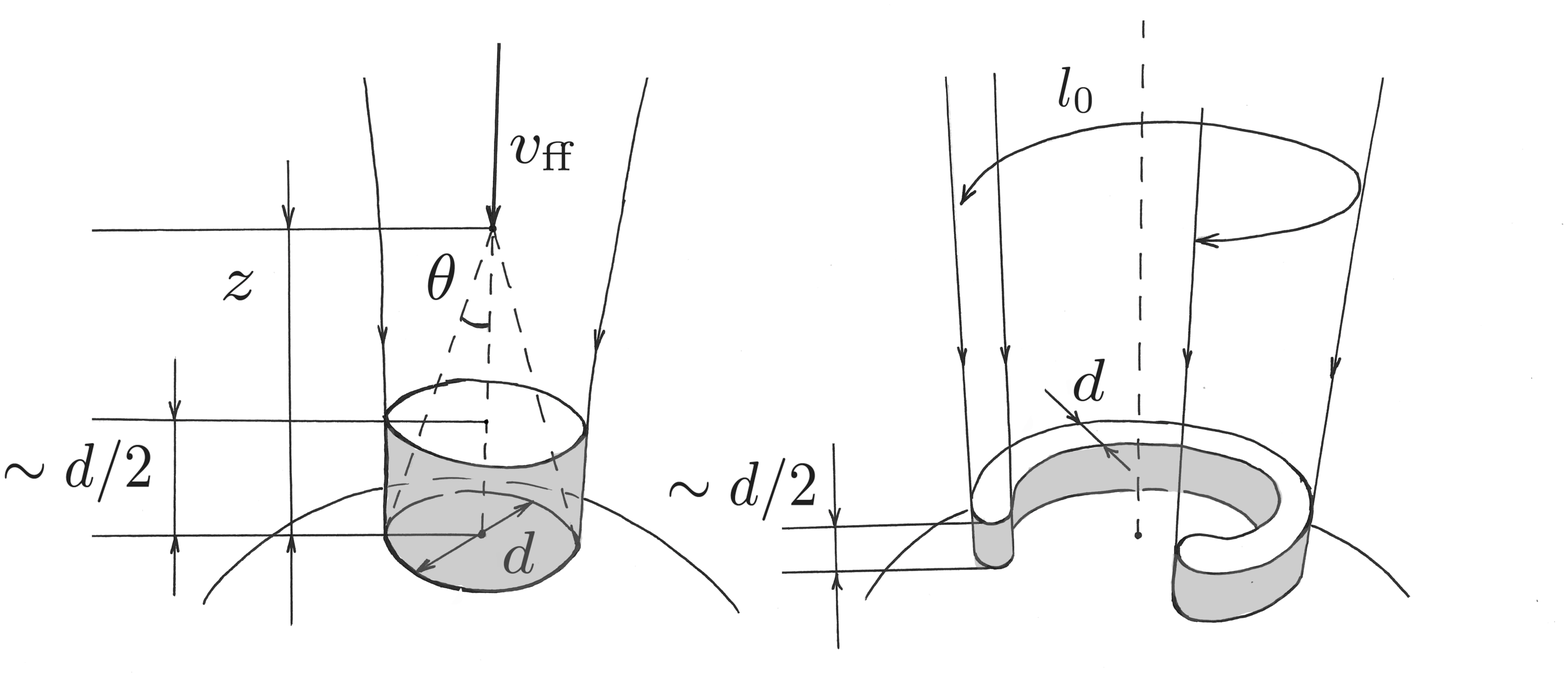} 
\caption{The infalling matter is decelerated by the radiation pressure force from the hotspot. 
  Photons are coming from the hotspot within the solid angle, which is defined by the distance 
  of a given point from the surface.Accretion via the wind  (left) and the disc (right) 
  leads to a different geometry of the accretion channel  and the hotspot
  shape, as well to the different structure of the braking region. 
  In any case the effective deceleration $g_{\rm R}$ drops
  rapidly with the distance from the surface $z$ and the radiation
  pressure decelerates the infalling matter effectively only in the layer
  $0<z\lesssim d/2$ \citep{BS1975,LS1982}.
  }
\label{fig_Model}
\end{figure*}

Following the ideas already discussed in the literature \citep{GS1973,MP1982},   
we compute here the critical luminosity accurately accounting for the first time 
for the influence of resonances in the Compton scattering cross-section, 
polarization, and the geometry of the accretion flow.
We base our calculations on  the physical model described by \citet{BS1976} 
where it is shown that the critical luminosity is not associated with the standard Eddington limit,
but should also account for braking the plasma infalling with high velocity 
above the NS surface.\footnote{A previous attempt to compute the critical
luminosity by \citet{Becker2012} was based on an erroneous assumption that 
the critical luminosity is associated with the Eddington limit. They have also neglected the possibility
of mixed polarization states and assumed that all photons are below the cyclotron energy 
neglecting thus strong resonances in the cross-section.}
Finally, we compare the obtained theoretical dependences of the critical 
luminosity on the magnetic field strength with the available 
observational data.

\section{Basic relations}
\label{sec:BasicIdeas}

The accretion column in XRPs arises as soon as the radiation pressure force $g_{\rm R}$
becomes high enough to stop the infalling matter from the free-fall velocity down to zero 
above the NS surface \citep{BS1975}. The necessary radiation pressure force is significantly larger than 
the Eddington radiation pressure force which balances the NS gravitational acceleration 
\be
    g_{\rm Edd} = \frac{GM}{R^2}(1-u)^{-1/2} . 
\ee  
Here $M$ and $R$ are NS mass and radius, $u=R_{\rm S}/R$ is the compactness parameter
and $R_{\rm S}=2GM/c^2$ is the NS Schwarzschild radius.

The radiative acceleration $g_{\rm R}$ necessary to stop the infalling matter can be evaluated with a simple
approach \citep{BS1975}. Let us assume that the accreting matter heats NS surface and forms a bright axisymmetric spot
of diameter $d$ ($d \ll R$) which radiates all the kinetic energy.
Ignoring any relativistic effects, one can find out the radiative acceleration at the distance $z$ above the spot:
\be \label{eq:gr1}
g_{\rm R} \approx \frac{2\pi}{c}\, \int_0^{ \infty} dE\, \int_{\mu_0(z)}^{ 1}\, \kappa (B,\mu, E)\, I(\mu, E) \, \mu\,d\mu,
\ee
where $\kappa$ is the opacity for the interaction process, 
$\mu = \cos\theta$ with $\theta$ being the angle measured from the radial direction 
and $\mu_0(z)=1/\sqrt{1+(d/2 z)^2}$ is defined by the angular size of the hotspot 
as seen from a given point above the NS surface (see Fig.~\ref{fig_Model}). 
Assuming 
isotropic specific intensity $I(\mu) \approx F/\pi$, where $F=L/2S$ is the hotspot bolometric flux, $S\approx \pi d^2/4$ is the spot 
area and  $L$ is the total XRP luminosity, one gets
\be\label{eq:gr2}
       g_{\rm R} \approx \frac{\kappa_{\rm eff}}{c}\, F \left[1-\mu_0^2(z)\right] \approx \frac{\kappa_{\rm eff}}{c}\, \frac{L}{2S}\,\frac{d^2}{d^2+4z^2} ,
\ee
where $\kappa_{\rm eff}$ is the effective opacity. 
The last term in equation (\ref{eq:gr2}) drops rapidly with the distance from the surface at $z > d/2$ \citep{BS1975} and $g_{\rm R}\propto (d/z)^{2}$.  As
a result the radiation pressure decelerates the matter effectively
only within a  layer $0<z\lesssim d/2$, where the radiation force is almost constant
\be
    g_{\rm R} \approx\frac{\kappa_{\rm eff}}{c}\,F.
\ee     
 
Taking into account the characteristic braking length $\sim d/2$ and assuming that the velocity of matter has to
drop from the free falling velocity $v_{\rm ff}\approx c\sqrt{R_{\rm S}/R}$ at $z \approx d/2$ down to zero at the surface, 
we can estimate the necessary deceleration: 
\be \label{eq:gg}
   g \approx \frac{v_{\rm ff}}{t_{\rm ff}} \approx \frac{v_{\rm ff}^2}{d},
\ee
where $t_{\rm ff}$ is a characteristic free-fall time.
Therefore, using an equality $g_{\rm R} = g$, we can find a hotspot flux $F^*$, which is sufficient to stop matter
by the radiation pressure  
\be
  F^{*} \approx \frac{c}{\kappa_{\rm eff}} \frac{v_{\rm ff}^2}{d} .  
\ee 
Corresponding critical luminosity for a circular spot is then \citep{BS1975, BS1976}\footnote{We define $Q=Q_x10^x$ in cgs units if 
not mentioned otherwise.}
\be\label{eq:cl}
     L^* \approx \frac{c}{\kappa_{\rm eff}}\pi d  \frac{GM}{R} 
     \approx  3.7\times 10^{36}\left(\frac{\kappa_{\rm T}}{\kappa_{\rm eff}}\right) 
\frac{d_5}{R_6} \mass\  \ergs,
\ee 
where $\kappa_{\rm T}\approx 0.34$\, cm$^2$\,g$^{-1}$ is the Thomson scattering opacity for 
solar composition material and $\mass = M/{\rm M}_\odot$. 

If accretion to the NS with dipole magnetic field  proceeds through the accretion disc, 
the matter is confined to a narrow wall of magnetic funnel 
and the hotspots have a shape of ring with a base length $l_0$ and width $d$ instead of circles (Fig.~\ref{fig_Model}). 
The estimation for the critical luminosity (\ref{eq:cl}) is also valid for this case if we replace $\pi d \rightarrow l_0$ 
\citep{BS1975, BS1976}. 

The approach described presented above assumes that all kinetic energy of the infalling matter is spent 
to heat the NS surface and accounts only for radiation from the hotspots.  
A realistic physical picture  is  more complicated: 
\begin{enumerate}
\item scattering on the infalling electrons is incoherent, so that 
the photons get a fraction of the electron momentum and energy;
\item photons are scattered predominantly downwards
heating the NS surface even more, and 
\item a fraction of them are scattered back towards the infalling matter and 
gets more energy and momentum. 
\end{enumerate}
As a result of this bulk Comptonization process \citep{BP1981a}, a radiation-dominated shock is formed  
\citep{ZR1967, BP1981b}, where matter is heated up and is decelerated from the supersonic infall velocity to the 
sub-sonic sedimentation velocity. The observed power-law-like spectra
of XRPs also could be formed due to the bulk Comptonization in the radiation-dominated shock \citep{LS1982, BW2007}. 
The optical depth of the radiation-dominated shock is about $\tau \sim 5-7$\, \citep{ZR1967}, and thus 
the considered above deceleration slab is optically thick (see next section).  Therefore, it is necessary to consider 
the radiation transfer accounting for thermal and bulk Comptonization and  the dynamics of the infalling plasma self-consistently. 
However, we expect that most of these complications are of little importance while 
the  major effects are coming from the energy dependence of   
the effective opacity which may change by orders of magnitude. 
Thus we assume that equation~(\ref{eq:cl}) is a good first approximation. 

The critical luminosity is defined by the mass and the radius of a star, geometry of the accretion flow near the
surface and the effective opacity. 
Interestingly, there is no explicit dependence of the critical luminosity on the thickness of the accretion channel. 
The effective opacity is determined by an effective cross section $\sigma_{\rm eff}$:
\be
\kappa_{\rm eff} =  \frac{\sigma_{\rm eff}}{\mu_{\rm e}\,m_{\rm p}},
\ee
where $\mu_{\rm e} \approx 1.18$ is the mean molecular weight per free electron
for the fully ionized plasma with the solar hydrogen-helium mixture.

The calculation of the effective cross-section is a key problem
here. In a general case of magnetized plasma, its value is determined
mainly by Compton scattering and cyclotron absorption \citep{HL2006}.
However, we are interested in NSs with very strong $B$-field
($\gtrsim 10^{12}$~G).  In such a fields the cyclotron
decay rate is quite high and an electron that absorbs a cyclotron
photon will be almost always de-excited by emitting another photon,
rather than be collisionally de-exited. As a result the resonant
scattering dominates over the true absorption (\citealt*{Bon1979}; \citealt{Her1982})
and the principal process in the interaction is Compton scattering.

Compton scattering cross-section for  high 
$B$-field depends strongly on photon energy \citep{Harding1986}, 
with large variations around the cyclotron harmonics. 
Thus, photons of different energies make very different contribution 
to the  radiation pressure force. 
Because in our calculations we do not compute the radiative transfer and 
the hotspot spectra self-consistently, we adopt a simple prescription 
for the photon spectrum that is fully determined by the
effective temperature $T_{\rm eff}$: 
\be\label{TeffBasic} 
\sigma_{\rm SB} T^4_{\rm eff}=F^*,
\ee 
where $\sigma_{\rm SB}$ is the Stefan-Boltzmann constant.

It should be noted that the luminosity and the temperature are given
here in the NS reference frame
and corresponding corrections for the observed luminosity, $L^{\infty}=(1-u)L$, 
and the cyclotron energy, $E^{\infty}_{\rm cycl}=(1-u)^{1/2}E_{\rm cycl}$, 
have to be made if one wants to compare  simulations with the  data.

\section{Hotspot area and effective temperature}
\label{SecHotSpot}

The shape and the area of the hotspots are determined by the structure of the 
accretion channel. 
The spot area is defined mainly by the interaction between the NS magnetosphere and
matter in the binary system. If a pulsar is fed from the accretion disc, the
accretion flow in the vicinity of a NS has geometry of a narrow
cylindrical ring. For the wind-fed pulsars, one can expect a completely filled funnel
cavity. The magnetospheric radius depends on the $B$-field strength and
structure, mass accretion rate and the way a pulsar is fed. 
It can be estimated with the following expression
\citep*{Lamb1973,fkr2002} 
\be\label{AlfvRad} 
R_{\rm m}=2.6\times 10^8
\Lambda\, m^{1/7}R_6^{10/7}B_{12}^{4/7}L_{37}^{-2/7} \ \mbox{cm},
\ee 
where $\Lambda$ is a constant which depends on the accretion flow
geometry: $\Lambda=1$ for the case of spherical or wind accretion 
(W-case; for example, in  Vela X-1),  
and $\Lambda<1$ for the case of accretion through the disc 
(D-case; expected e.g. in Her X-1, GX 304--1, V 0332+53,
4U 0115+63), with  $\Lambda=0.5$ being a commonly
used value \citep{GL1978, GL1979}. 
The W- and D-accretion scenarios give quite different
predictions for the spot area because of the different structure of
the accretion channel near the NS surface.  At the same time
we can consider these two scenarios as limiting cases because the
accretion process in many system may partly proceed both
ways.

In the W-case scenario, the hotspot area can be expressed as 
\begin{eqnarray} \label{eq:S_W}
&&S_W \approx  \pi d^2/4 \approx \pi R^3 R_{\rm m}^{-1} \\ \nonumber  && \approx 1.3 \times 10^{10}
\,\Lambda^{-1}\, m^{-1/7}\,  R^{11/7}_6
\, B^{-4/7}_{12} \, L^{2/7}_{37} \quad \mbox{cm}^2.  
\end{eqnarray}
under assumption of the dipole configuration of the $B$-field. 
Using equation (\ref{TeffBasic}) we then immediately get
the effective temperature:
\be \label{eq:T_W}
T^{W}_{\rm eff}=4.5\,\Lambda^{1/4}
B_{12}^{1/7} \, L_{37}^{5/28} m^{1/28}\,
R_6^{-11/28}\quad \mbox{keV}.  
\ee 
Expressions (\ref{eq:S_W}) and (\ref{eq:T_W}) give us 
the maximum hotspot area and, correspondingly, the minimum effective
temperature for the fixed mass accretion rate. 

In the D-case, matter comes closer to the NS 
(see equation (\ref{AlfvRad})), the plasma is confined to a narrow
wall of the magnetic funnel. The thickness of the accretion channel  depends on
the penetration depth of the accretion disc into the NS magnetosphere
\citep{Lai2014}, which is expected to be of the order of
$\delta\approx 2H$, where $H$ is a disc scale-height at the inner edge
\citep{GL1978,GL1979}. 
We evaluate $H$  using the \citet{ShSu1973} model, 
slightly modified how the vertical structure is averaged and using the correct Kramer opacity \citep*{Sul2007}.
The radius of magnetosphere for adopted parameters is situated 
in the so-called $C$-zone of accretion disc, where gas pressure and Kramer opacity dominate.
The boundary of this zone for adopted parameters is situated at \citep{Sul2007}
\be
      r > r_{BC} \approx 5.5 \times 10^7 \, L_{37}^{2/7}\, R_6^{2/7}\, m^{1/21} \quad \mbox{cm}.
\ee
A relative disc scale-height at radius $r$ for this zone is
\beq\label{SS_H}
\frac{H}{r}\approx 0.08 \,\alpha^{-1/10}\, L_{37}^{3/20}\, m^{-21/40}\, R_6^{3/20}\,r_8^{1/8},
\eeq
where  $\alpha < 1$ is a dimensionless viscosity parameter \citep{ShSu1973}.
Substituting  $R_{\rm m}$ from equation  (\ref{AlfvRad}) to equation (\ref{SS_H}) instead of $r$, we get
\beq
\frac{H_{\rm m}}{R_{\rm m}}
= 0.1\, \alpha^{-1/10}\,\Lambda^{1/8}
\, m^{-71/140}\, R_6^{23/70}\,B_{12}^{1/14}\,L_{37}^{4/35}.
\eeq
Then, the area of a single hotspot, which
has a shape of a closed ring on the stellar surface, is
\begin{eqnarray}
&&S_{D}=l_0 d \approx  2\pi
\frac{R^3}{R_{\rm m}}\frac{H_{\rm m}}{R_{\rm m}} \approx S_{W} \frac{2H_{\rm m}}{R_{\rm m}}\\ \nonumber
&& \approx 3 \times 10^9\, \Lambda^{-7/8}\, m^{-13/20}\, R_6^{19/10}\, B_{12}^{-1/2}\, L_{37}^{2/5}\quad     \mbox{cm}^2.  
\end{eqnarray}
The corresponding effective temperature  is
\beq
T_{\rm  eff}^{D} = 6.6 \, \Lambda^{7/32}\, m^{13/80}\, R_6^{-19/40}\, B_{12}^{1/8}\, L_{37}^{3/20} \quad \mbox{keV}.
\eeq
It is interesting that the obtained $T_{\rm  eff}^{D}$ is close to the temperature of the hot electrons 
$T_{\rm e} \approx 5.1$\, keV, when the observed spectrum of X-ray pulsar GX 301--2 was fitted by {\sc comptt} 
model \citep{Dor.etal2010}.

\begin{figure*}
\centering 
\includegraphics[width=8cm]{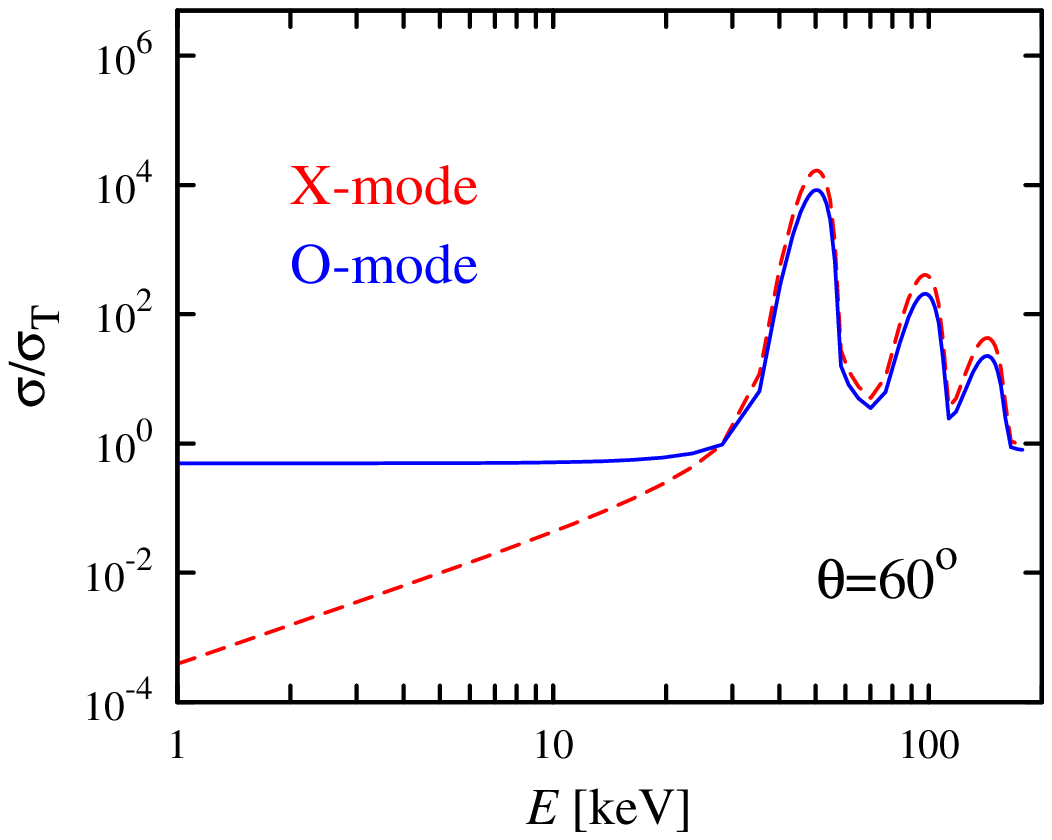} 
\includegraphics[width=8cm]{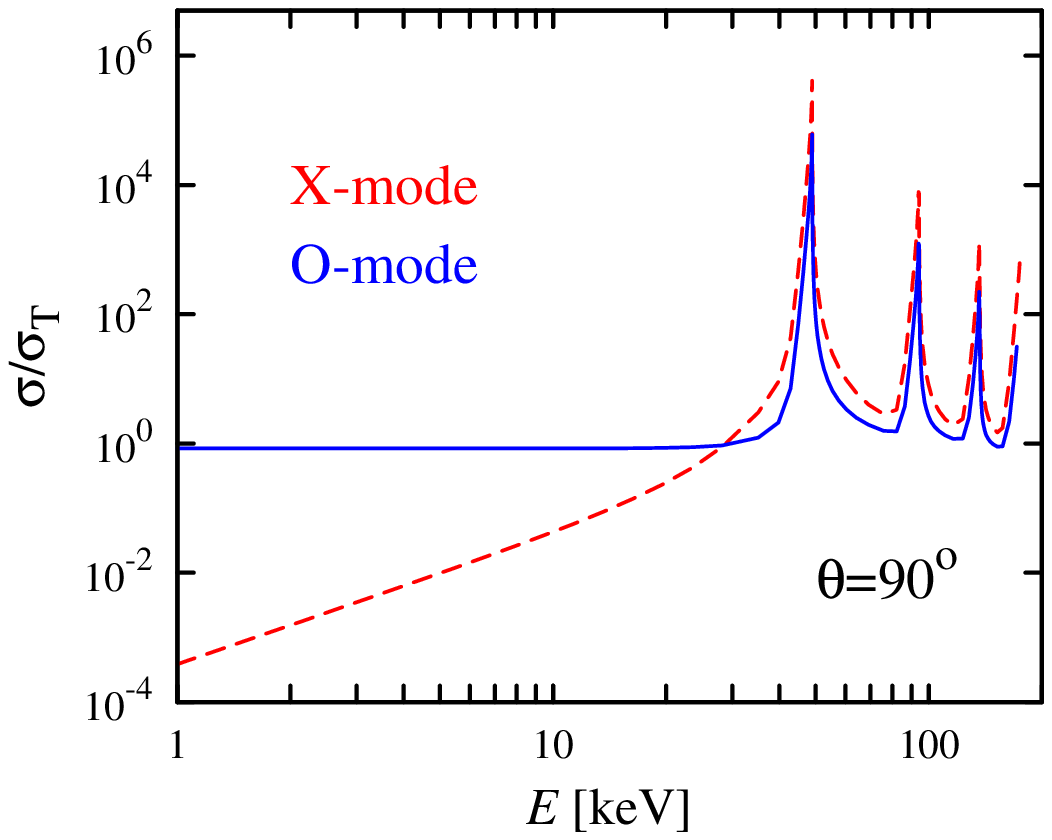} 
\caption{Total Compton scattering cross-section dependence on photon energy, 
polarization state (for X- and O-modes) and and the angle $\theta$ 
between the initial momentum direction and the $B$-field. 
The resonant features are broadened due to thermal motion of the electrons. Their positions 
depend on $\theta$.  
Here the magnetic field strength is  $b=0.1$ (i.e. $B_{12}\simeq 4.4$) and the electron temperature
$T_{\rm e}=5$ keV.
}
\label{pic:cs} 
\end{figure*}

If the magnetic dipole is inclined with respect to the orbital plane, 
the expressions for the hotspot areas would be different because the spot would 
have a shape of an open ring. 
It is reasonable then to use an additional
parameter $l_0/l$, which shows what part of the full ring length $l$
is exposed to accretion. We use this parameter further and analyse its influence on the final results.

\subsection{Thomson optical thickness}

The infalling plasma is stopped by the radiation force at the distance
which is comparable to the thickness of the accretion channel $d$
\citep{BS1976}. Under the assumption of linear velocity decrease to zero
value over the braking distance, the Thomson optical thickness of this
layer is 
\begin{eqnarray} \label{eq_taut}
\tau_{\rm T} & \gtrsim & \kappa_{\rm T}\,\frac{\dot M}{2S_{D}\, v_{\rm ff}}\, d  \approx \kappa_{\rm T}\,\frac{\dot M}{4\pi r_{0}\, v_{\rm ff}}\\ \nonumber
&&\approx 5\, \Lambda^{1/2}\,L^{6/7}_{37}\,B^{2/7}_{12}\, m^{-10/7}\, R_6^{10/14}, 
\end{eqnarray} 
where $\dot M \approx  1.3\times 10^{17}\, L_{37}\, m^{-1}\, R_6$\,\,g\,s$^{-1}$ is the mass accretion rate.
Thus, the plasma is optically thick for the luminosity and the range of 
magnetic field strengths  
which are of interest here. 
The actual optical depth can be much smaller than $\tau_{\rm T}$ 
in the case of ultra-strong magnetic field, when photon energies 
are far below the cyclotron energy and the scattering cross-section is quite small.
On the other hand, it can be much larger than $\tau_{\rm T}$ 
if cyclotron resonances occur close to  the peak of the spectral energy distribution.

\section{Compton scattering cross-section}
\label{secCS}

\subsection{Calculations of Compton scattering cross-section}
\label{cross_sec_gen}

The cross-section of Compton scattering in strong magnetic field
differs substantially  from the cross-section of this process in low $B$-field. 
It depends strongly on the photon energy
$E$, polarization mode $j$ ($j=1$ for the extraordinary mode --
"X-mode", and $j=2$ for the ordinary mode -- "O-mode"), the angle $\theta$
between the $B$-field direction and the photon momentum, the strength of the  $B$-field 
and the electron temperature $T_{\rm e}$. 
Resonances in the photon-electron interaction lead to extremely high values of the 
cross-section  around the cyclotron frequency and its harmonics
(Fig.~\ref{pic:cs}).  The exact positions of the resonances depend on
the field strength and on the photon momentum direction:
\begin {equation} 
\frac{E^{(n)}_{\rm res}(B)}{m_{\rm e}c^2 } \! =\!  
\left\{  \begin{array}{ll}
\strut\displaystyle \! \! 
\frac{\sqrt{1+2nb\sin^2\theta}-1}{\sin^2\theta}, & \mbox{for}\ \theta\neq0, \  n=1,2,..., \\
\! \!  b ,                                             & \mbox{for}\ \theta=0, 
 \end{array} \right. 
\end{equation} 
where $b\equiv B/B_{\rm cr}$ is the $B$-field strength in units of the
critical field strength $B_{\rm cr}= m_{\rm e}^2 c^3/e\hbar =
4.412\times 10^{13}$ G and $m_{\rm e}$ is the electron mass.

We calculate Compton scattering cross-section using 
the second order perturbation theory in quantum electrodynamics. 
Such calculations were already done by \citet{PShIa1980}, \citet{Harding1986} and \citet{Harding1991} 
(see A. Mushtukov et al., in prep., for the details) 
and the formalism was discussed partly by \citet{Mushtukov2012}.
The electron temperature $T_{\rm e}$ noticeably affects the cross-section
near the resonance energies by thermal broadening of the peaks. 
Because electrons in strong $B$-field move mostly along the field lines, 
the broadening  depends also on the angle between photon momentum and the field
direction. Thermal broadening has its maximum for the photons
which propagate along the field and minimum for photons
moving in the perpendicular direction because only the relativistic transverse
Doppler effect operates in this case.  
The scattering cross-section by an ensemble of electrons described by the 
distribution function over the longitudinal momentum $f(Z,T_{\rm e})$ 
(normalized to unity $\int_{-\infty}^{\infty}f(Z,T_{\rm e})\d Z=1$) is  \citep{Harding1991}: 
\beq
\sigma(E,\mu,T_{\rm e})=\int\limits_{-\infty}^{\infty}\d Z\,
f(Z,T_{\rm e})\sigma_{\rm R}(E_{\rm R},\mu_{\rm R})\ \gamma(1-\beta\mu), 
\eeq 
where $\sigma_{\rm R}(E_{\rm R},\mu_{\rm R})$ is the cross-section for electrons at rest 
(indicated by subscript R). Here 
$\mu=\cos\theta$ and $\mu_{\rm R}$ are related by the relativistic aberration formula,
$\beta=v/c$ is dimensionless electron velocity
corresponding to the dimensionless electron momentum $Z=\beta\gamma$ and
$\gamma=(1-\beta^2)^{-1/2}$ is the  Lorentz factor.

\begin{figure*}
\centering 
\includegraphics[height=6cm,angle=0]{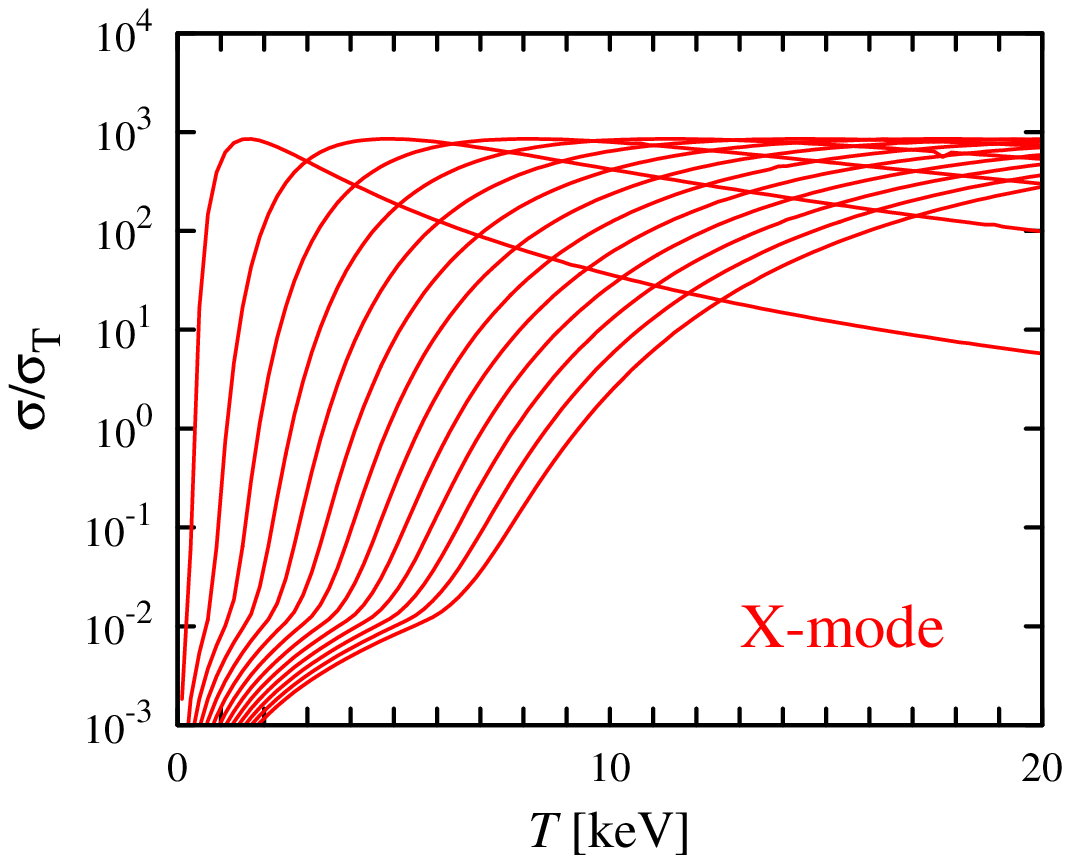} 
\includegraphics[height=6cm,angle=0]{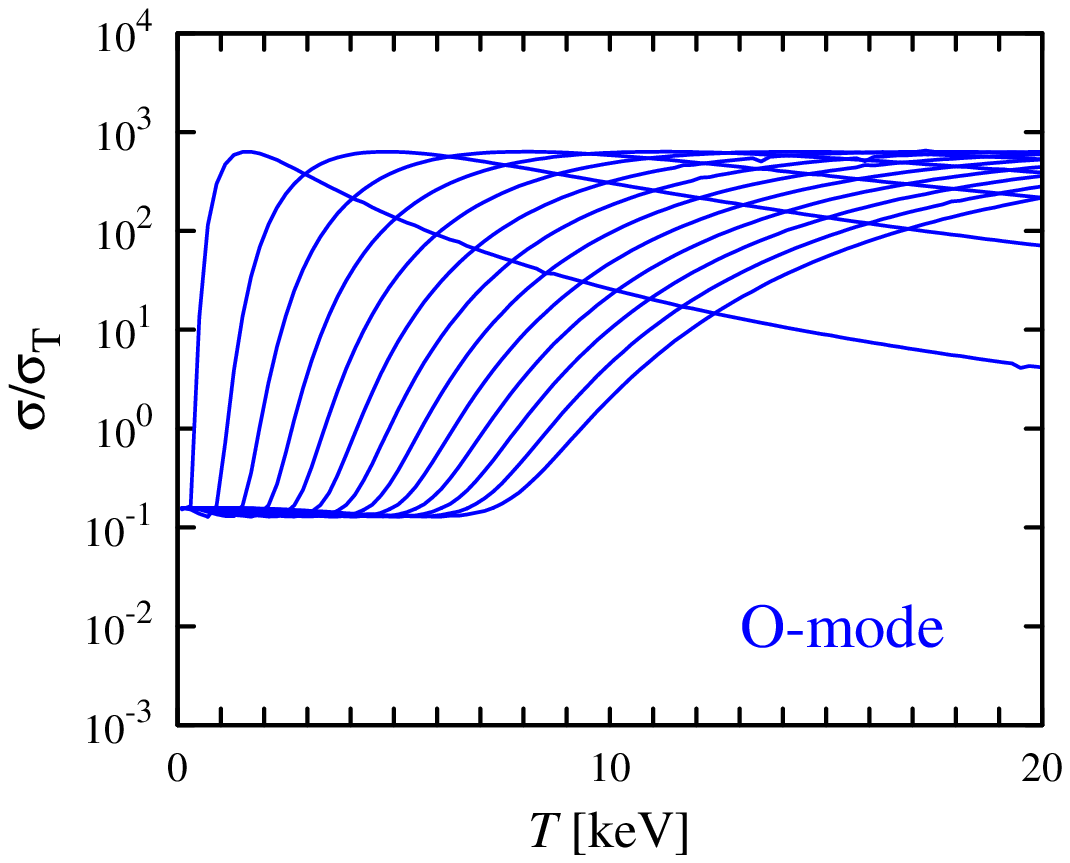} 
\caption{The mean cross-section for photons of X- and
  O- polarization modes as a function of the blackbody temperature (see equation~(\ref{eq:cs_simp})).
  The different lines correspond to various magnetic field strength
  starting from $b=0.02$ up to $b=0.5$ with a step $\Delta b=0.04$ (from left to right).
  The electron temperature is fixed at $T_{\rm e}=1$~keV.}
\label{pic:cs_simp} 
\end{figure*}

A good lower estimate for the electron temperature $T_{\rm e}$ of the accretion flow 
comes from the fact that for luminous pulsars X-ray radiation keeps the gas at the Compton temperature. 
For a typical XRP spectrum with photon index $\Gamma\approx 1$--1.5 and a cutoff at 20--30 keV, 
the Compton temperature lies in the interval 1--5 keV.
The upper limit on the temperature comes from the cutoff energy and gives  $T_{\rm e}\sim 10$~keV.
In further calculations we assume $T_{\rm e}=1$~keV 
and discuss the effects of possible deviation from that value.

\subsection{Effective cross-section}

The effective cross-section defines the radiation pressure, and,
hence, the critical luminosity for a given magnetic field strength.
It depends on the photons distribution over energy, directions of
propagation and polarization states.

The expression for radiation force could be written in the following way \citep[see e.g. equation (2.50) in][]{PSS1983}:
\beq\label{F_R:gen}
 \strut\displaystyle 
g_{\rm R}&=&\frac{1}{c}\sum\limits_{j,n_\ff}\int\limits_{0}^{\infty}\d E_\ii
\int\limits_{(4\pi)}\d\Omega_\ii
\int\limits_{0}^{\infty}\d E_\ff\int\limits_{(4\pi)}\d\Omega_\ff
\\
 \strut\displaystyle
&\times& \frac{\d\sigma_{j}(E_\ii,\mu_\ii|E_\ff,\mu_\ff,n_\ff,T_{\rm e})}{\d E_\ff\d\Omega_\ff} 
I_{j}(E_\ii,\mu_\ii)\left(\mu_\ii-\frac{E_\ff}{E_\ii}\mu_\ff\right), \nonumber
\eeq
where $\d\sigma_{j}/\d E_\ff\d\Omega_\ff$ is the
differential cross section, indexes "$\ii$" and "$\ff$" denote the initial and
final particle conditions, $I_{j}(E_\ii,\mu_\ii)$ is the intensity
corresponding to a given photon polarization state $j$ (X- or
O-mode), energy $E_\ii$ and direction $\mu_\ii$.  The difference
$\left(\mu_\ii-\mu_\ff{E_\ff}/{E_\ii}\right)$ defines the recoil effect in
each scattering event. The summation is made over the final Landau
level numbers $n_\ff$ and photon polarization states $j$.  The total
scattering cross section for a given polarization state is
\be
\sigma_{j}(E_\ii,\mu_\ii,T_{\rm e})\!  =\!  \sum_{n_\ff}\int\limits_{0}^{\infty} \!   \d E_\ff \! \! \! \!  \int\limits_{(4\pi)}\! \!\!  \d\Omega_\ff
\frac{\d\sigma_{j}(E_\ii,\mu_\ii|E_\ff,\mu_\ff,n_\ff,T_{\rm e})}{\d E_\ff\d\Omega_\ff}.
\ee

If the photon redistribution is symmetric relative to the plane
perpendicular to the initial photon momentum (it is a reasonable
assumption for the relatively low-energy photons $E\ll m_{\rm e}c^2$), 
then each photon on average transfer its own momentum to
the electron and the expression for the radiation force
(\ref{F_R:gen}) can be simplified: 
\beq
g_{\rm R}=\frac{1}{c}\sum\limits_{j}\int\limits_{0}^{\infty}\d E
\int\limits_{(2\pi)}\d\Omega\ \sigma_{j}(E,\mu,T_{\rm e}) \ I_{j}(E,\mu)\ \mu.
\eeq 
In the following we use this simplified expression because photons emitted from
the hotspots are not expected to have energies higher than $\sim20$~keV.

For optically thin plasma falling onto the NS 
surface, the hotspot radiation spectrum does not change 
and the expression for the radiation force in the
plasma reference frame takes the form 
\beq\label{F_R:app}
g_{\rm R}\approx\frac{\pi}{c}\sum\limits_{j}\int\limits_{0}^{\infty}\d E
\int\limits_{\mu^*}^{1}\d\mu\,\mu\ \sigma_{j}(E,\mu,T_{\rm e})\ B_{\rm E}(T(\mu)),
\eeq 
where $B_{\rm E} (T)$ is the Planck function, $T(\mu)=T_{\rm  eff}\gamma(1+\beta\mu)$ and
$\mu^*=(\mu_{0}+\beta)/(1+\beta\mu_{0})$, $\mu_{0}$ is cosine
of the maximum polar angle from which spot radiation is coming.  
Equation~(\ref{F_R:app}) is written for the axisymmetric case 
assuming blackbody radiation. 
The light aberration and the transformation of the radiation field due to the Lorentz
transformation are taken into account. Because by definition 
$g_{\rm R}=\sigma_{\rm eff}F/c$, the effective cross section in this case is 
\be\label{eq:cs_simp}
\sigma^{(1)}_{\rm eff}=\frac{\sum\limits_{j}\int\limits_{0}^{\infty}\d E
  \int\limits_{\mu^*}^{1}\d\mu\,\mu\ \sigma_{j}(E,\mu,T_{\rm e})\ B_{\rm E}(T(\mu))}
      {\sum\limits_{j}\int\limits_{0}^{\infty}\d E
        \int\limits_{0}^{1}\d\mu\,\mu\ B_{\rm E}(T_{\rm eff})}.  
\ee
The results of calculations are shown in Fig.~\ref{pic:cs_simp}.

\begin{figure*}
\centering 
\includegraphics[height=6cm,angle=0]{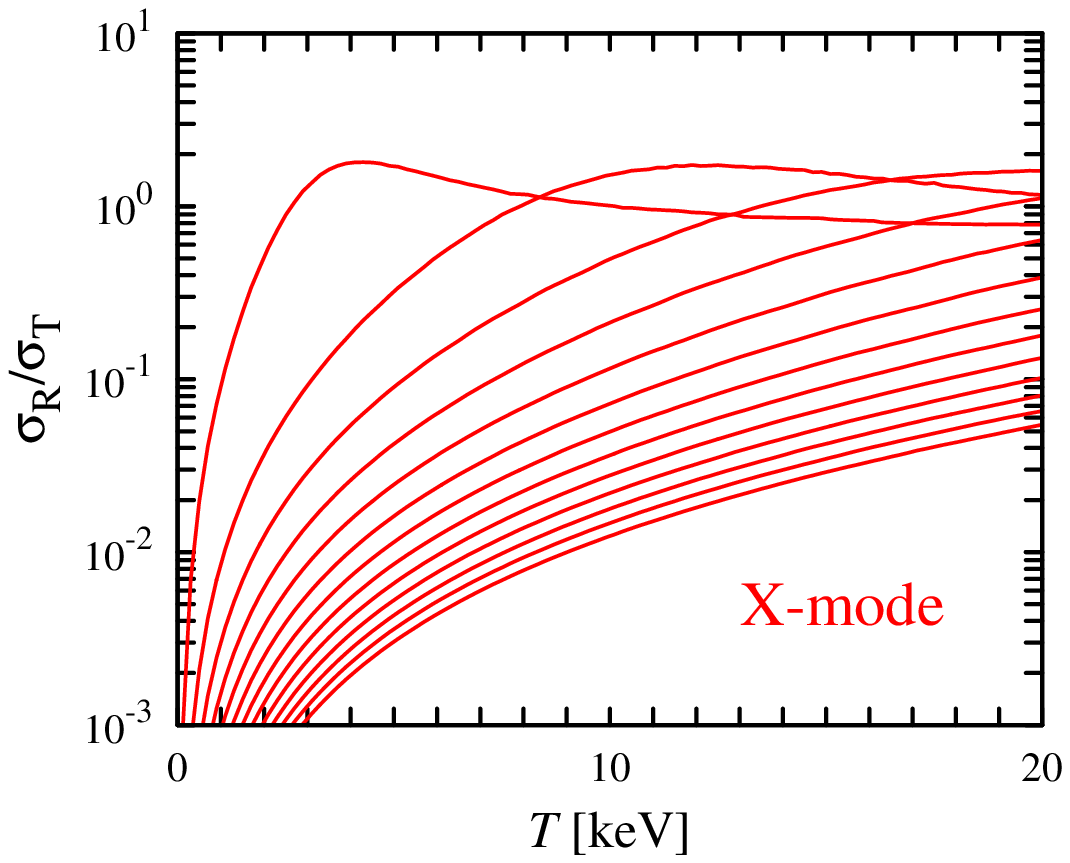} 
\includegraphics[height=6cm,angle=0]{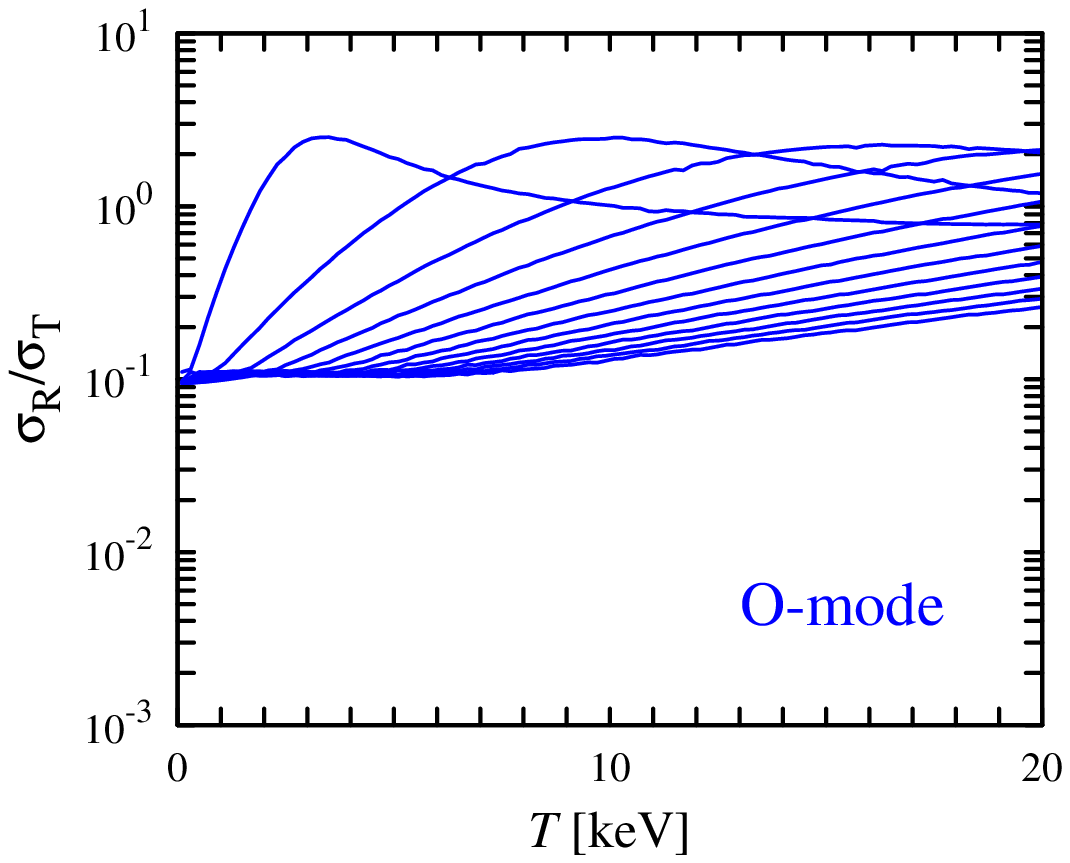} 
\caption{The Rosseland mean cross-section for photons of X- and
  O- polarization modes as a function of the blackbody temperature.
  The different lines correspond to various magnetic field strength
  starting from $b=0.02$ up to $b=0.5$ with a step $\Delta b=0.04$ (from left to right).
  The electron temperature is fixed at $T_{\rm e}=1$~keV.}
\label{pic:cs_Ross} 
\end{figure*}

On the other hand, in the case of the optically thick plasma, the radiation
field could be modified significantly and the situation is much more
complicated. In general, it is necessary to calculate accurately the
radiation transfer problem together with the structure of a
radiation-dominated shock  near the surface. 
However, it is also possible to get the approximate effective cross-section 
using the Rosseland approximation \citep{Putten2013}. For the 
angle-independent cross-section, the
Rosseland mean value has a well known form 
\be
\frac{1}{\sigma_R}=\frac{\displaystyle\int_{0}^{\infty} \frac{\d B_E(T)}{\d
    T}\frac{1}{\sigma(E)}\d E} {\displaystyle\int_{0}^{\infty}
    \frac{\d B_{\rm E}(T)} {\d T}\d E}.  
 \ee 
For the angle-dependent cross-section the
expression can be generalized: 
\be\label{eq:ross_ang} 
\frac{1}{\overline{\sigma}_{j}}=
\frac{\displaystyle\int\limits_{0}^{\infty} 
   \frac{\d B_{\rm E}(T)} {\d T}\d E  
\int\limits_{0}^{1}\d\mu\,3\mu^2 
  \frac{1}{\sigma_{j}(E,\mu ,T_{\rm e})}}
  {\displaystyle\int_{0}^{\infty}
    \frac{\d B_{\rm E}(T)} {\d T}\d E}. 
  \ee 
The  Rosseland cross-section as a function of the $B$-field strength and the temperature 
is shown in Fig.~\ref{pic:cs_Ross} for both polarizations.    
We note that the cross section is significantly smaller than that for the optically thin case.

The hotspot radiation is interacting with the moving plasma inside the radiation shock region
and braking it. The spectra in the star reference frame and in the
moving electron reference frame are different due to the Doppler
shift and relativistic aberration.  Moreover the interaction
between infalling plasma and the hotspot radiation changes the spectrum. As
a result the characteristic photon energy is a bit higher than it is
expected from the obtained effective temperature (Section~\ref{SecHotSpot}).  
The problem could be solved approximately with the
correction of the effective temperature in equation (\ref{eq:ross_ang}).  Under
the assumption of free photon escape from the accretion channel walls
and free supersonic gas infall down to the shock front, the velocity
profile inside the shock region is $v(z)=v_{\rm ff}(1-\exp[-z/d])$,
where $z$ is the height of a given point \citep{LS1982}. The photon energy shift in the electron
reference frame depends on the electron velocity and the angle between photon and electron momenta. 
Taking into account the electron velocity profile in the shock region and distribution of photons 
over momentum direction we estimate that the radiation temperature in the plasma frame 
is larger than the effective temperature by a factor $1<\zeta\lesssim 1.5$, i.e. 
we should use $T=\zeta T_{\rm eff}$ in equation (\ref{eq:ross_ang}).
The initial hot spot spectrum is not well known and might differ from the blackbody spectrum. The corresponding measure
of the mean photon energy is corrected by a factor of 0.5--1, with the lower value corresponding to the bremsstrahlung
and the upper value to the Planck spectrum. This difference can be also taken into account by $\zeta$-coefficient. 
In our primary calculations we use $\zeta=1.5$ and discuss an influence of the coefficient on the final results in Section \ref{Sec:CritLum}.

Rosseland mean value (\ref{eq:ross_ang}) is written for the case of fixed photon polarization state.
In the case of mixed polarization the effective cross-section can be expressed as
\be
1/\sigma^{(2)}_{\rm eff}=\eta/\overline{\sigma}_{1}+(1-\eta)/\overline{\sigma}_{2},
\ee
where $\eta$ is a fraction of radiation in the X-mode.
This equation is written under assumption that the photon fraction of each polarization
does not depend on photon energy. In reality the problem could be more complicated and
it can be a function of photon energy and even of direction, but 
for the simple estimations it is reasonable to assume that photons of different polarizations are
mixed in some proportion. We use $\eta$ as a parameter in our calculations. 
Because the optical depth of the shock is large (see equation~(\ref{eq_taut})), 
it makes more sense to use the Rosseland mean opacity for calculation of the critical luminosity.

\begin{figure*}
\centering 
\includegraphics[width=8.5cm]{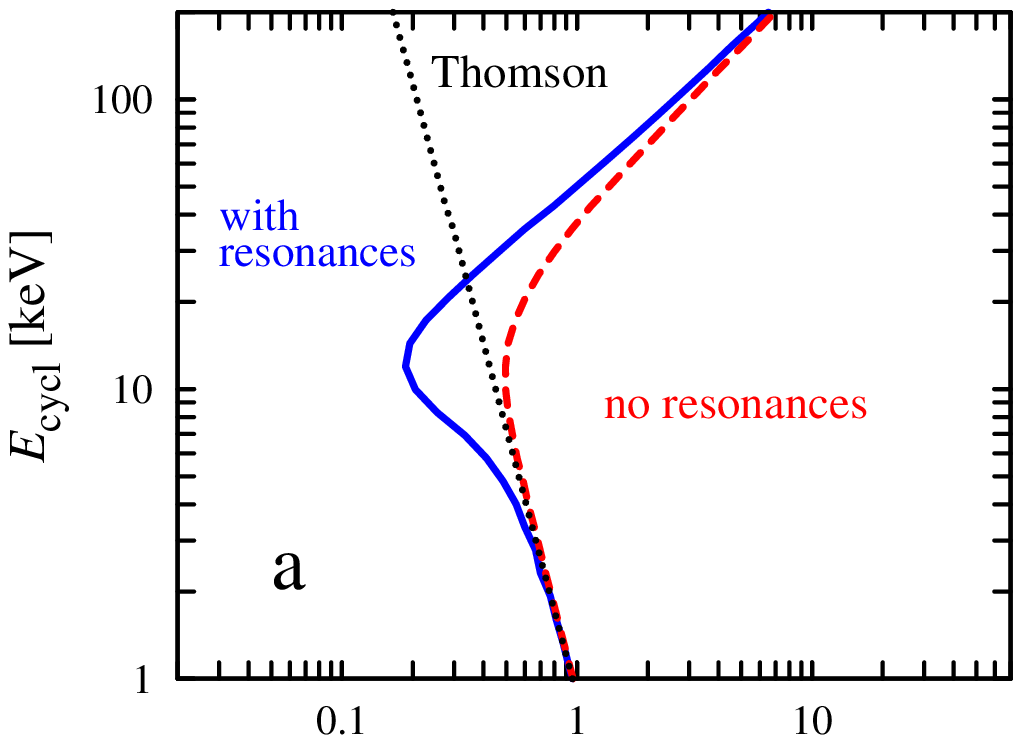} 
\includegraphics[width=8.5cm]{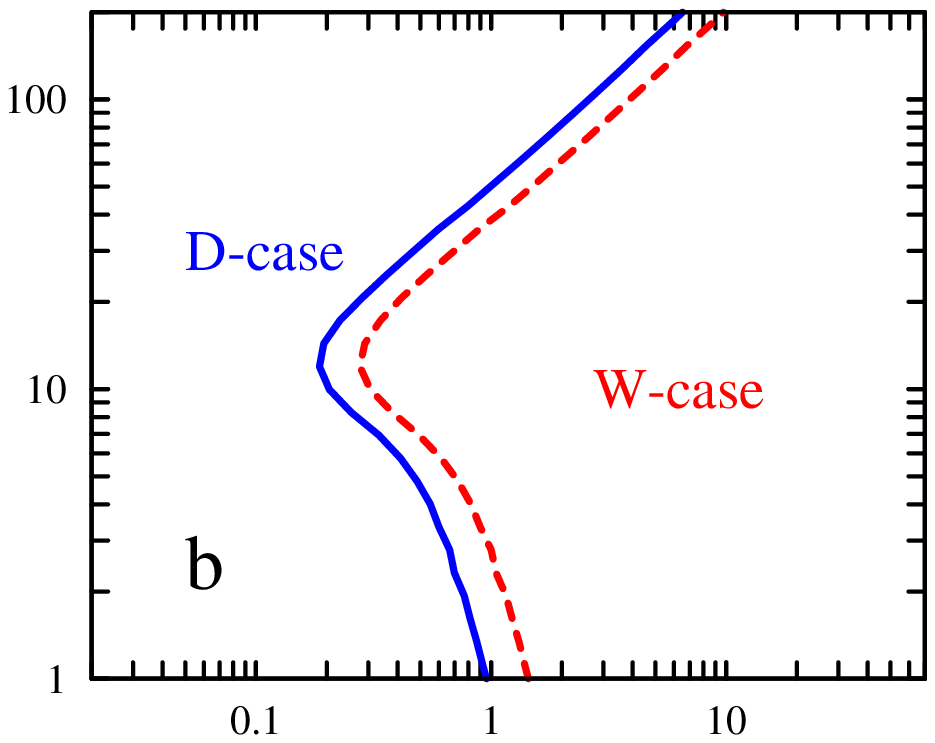} 
\includegraphics[width=8.5cm]{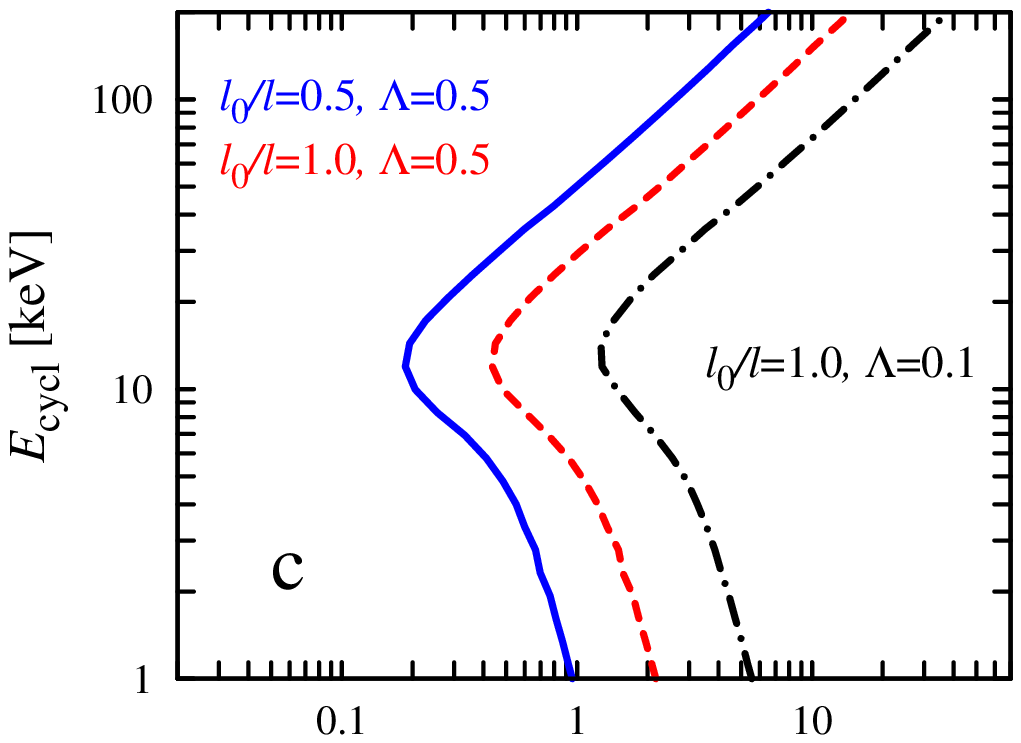} 
\includegraphics[width=8.5cm]{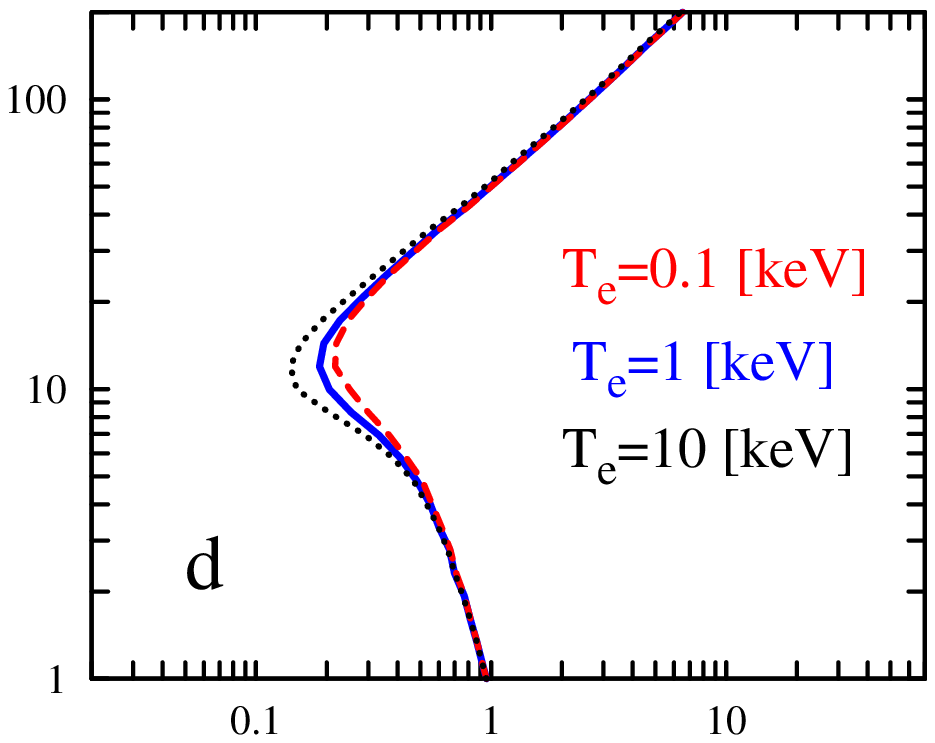} 
\includegraphics[width=8.5cm]{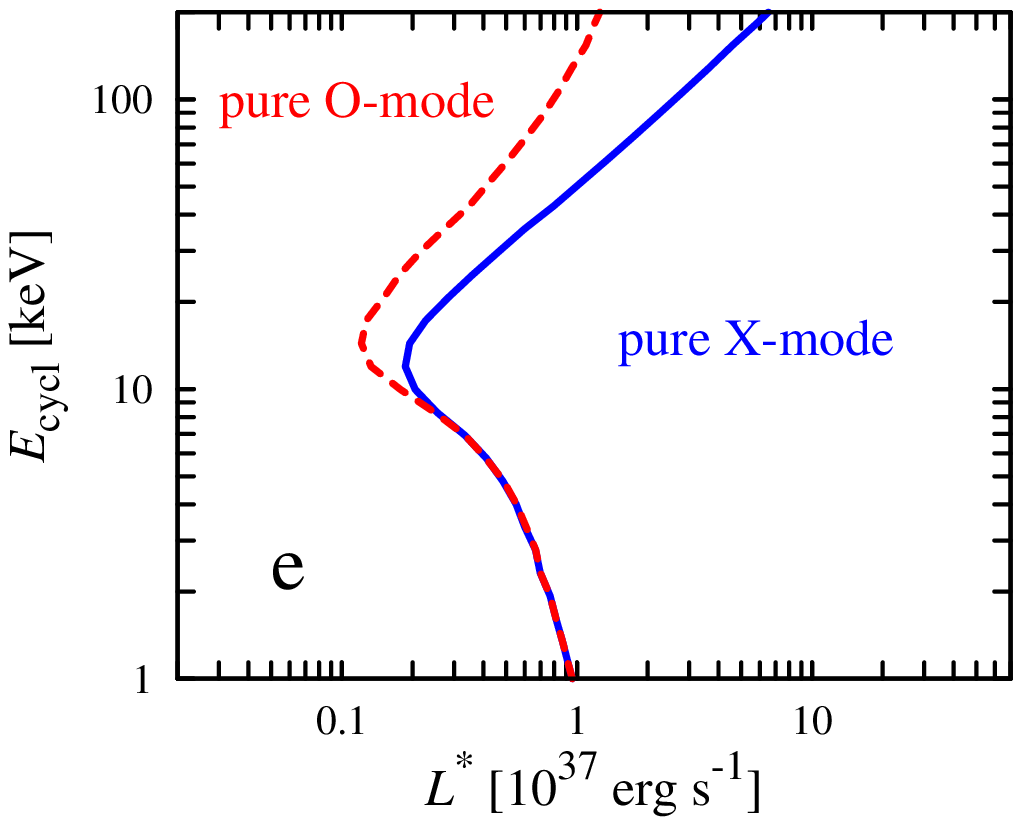} 
\includegraphics[width=8.5cm]{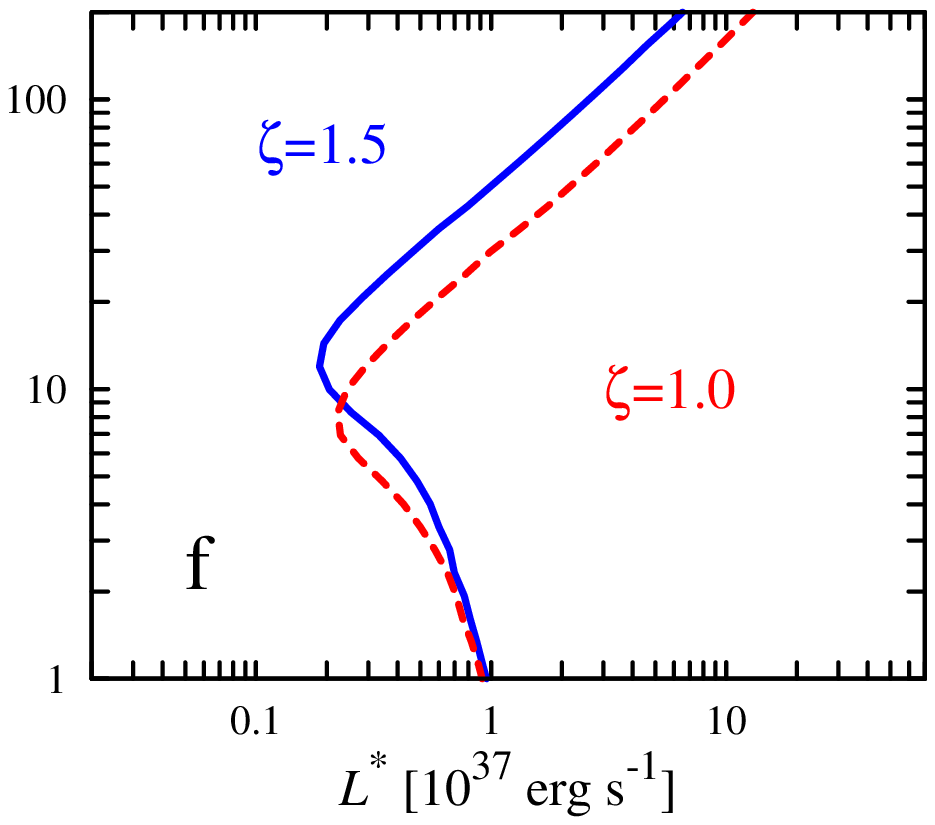} 
\caption{Dependence of the critical luminosity on the magnetic field strength expressed as 
an inverse relation of the cyclotron energy on luminosity. The fiducial case is given by solid blue line
and corresponds to disc-case with the following set of parameters: 
$m=1.4$, $R_6=1$, $l_{0}/l=0.5$, $\Lambda=0.5$, $T_{\rm e}=1$ keV,
$\zeta=1.5$, pure X-mode polarization is also assumed.
 (a) The effect of the resonant scattering and deviation of the cross-section from  the Thomson value. 
At low $B$ (i.e. low cyclotron energies), 
the critical luminosity is similar to that  for Thomson cross-section (shown by black dotted line). 
Resonances reduce the critical luminosity, when the cyclotron line energy is close 
to the typical photon energy from the hotspot.
(b) Dependence on the accretion flow structure. 
The wind case predicts slightly higher critical luminosity, 
mostly because of the larger hotspot size that leads to smaller temperature and larger breaking distance.
(c) Influence of the geometry of the accretion channel and the size of the magnetospheric radius.  
(d) Electron temperature in the accretion flow affects the resonance width and is important only for the case
when most of the photons come inside of the resonance.
(e) Dependence on polarisation. In high magnetic field 
the scattering cross-section for different photon polarizations is vastly different.  
(f) Effects of effective temperature change due to the motion of optically thick plasma. 
}
\label{pic:param}
\end{figure*}

\section{Results}
\label{Sec:Res}

\subsection{Critical luminosity}
\label{Sec:CritLum}

The main parameters defining the effective cross-section are the magnetic field strength (i.e. cyclotron energy $E_{\rm cycl}$), 
the effective temperature $T_{\rm eff}$ and the polarization mixture $\eta$. 
Once  the effective (Rosseland) cross-section is obtained, we can compute the critical luminosity. 
The results obviously depend on the radiation field structure. In
order to estimate the critical luminosity approximately we assume the
blackbody spectrum with the effective temperature which is defined by
the mass accretion rate and the hotspot area (see Section \ref{SecHotSpot}).
The   critical luminosity also depends on the NS mass and radius and accretion flow geometry. 
The expression for the critical luminosity is not linear: the effective cross section in the right hand side 
of equation (\ref{eq:cl}) is defined by the effective temperature and therefore by the hot spot area, which depends 
on the mass accretion rate or luminosity. Thus, the expression for the luminosity could not be used directly. 
We solve the problem in an iterative way. We start with fixed magnetic field strength and some reasonable luminosity value ($L_{37}=1$). 
They give us the spot area and the effective cross section. Then we compute new luminosity value using equation (\ref{eq:cl}) 
which gives the new effective cross section. This procedure continues until the difference between new and previous 
luminosities is sufficiently small. In the end we have self-consistent values for the critical luminosity 
and the effective cross section for a given $B$-field strength.

\begin{figure*}
\centering 
\includegraphics[height=6cm,angle=0]{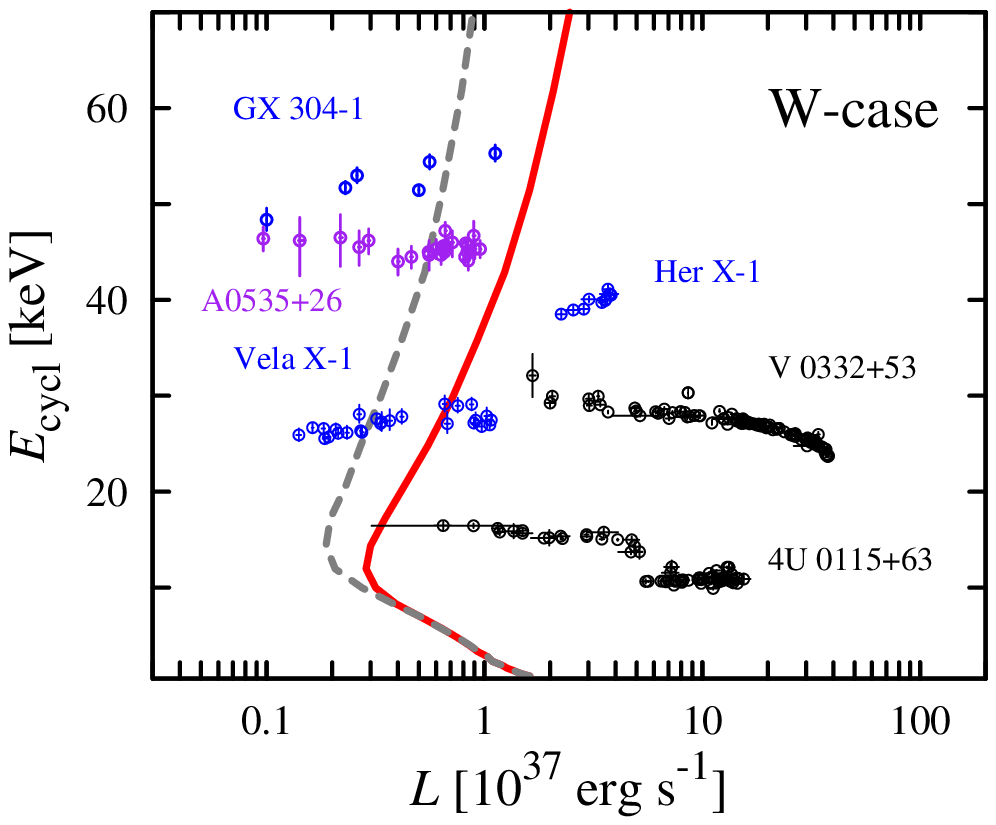} 
\includegraphics[height=6cm,angle=0]{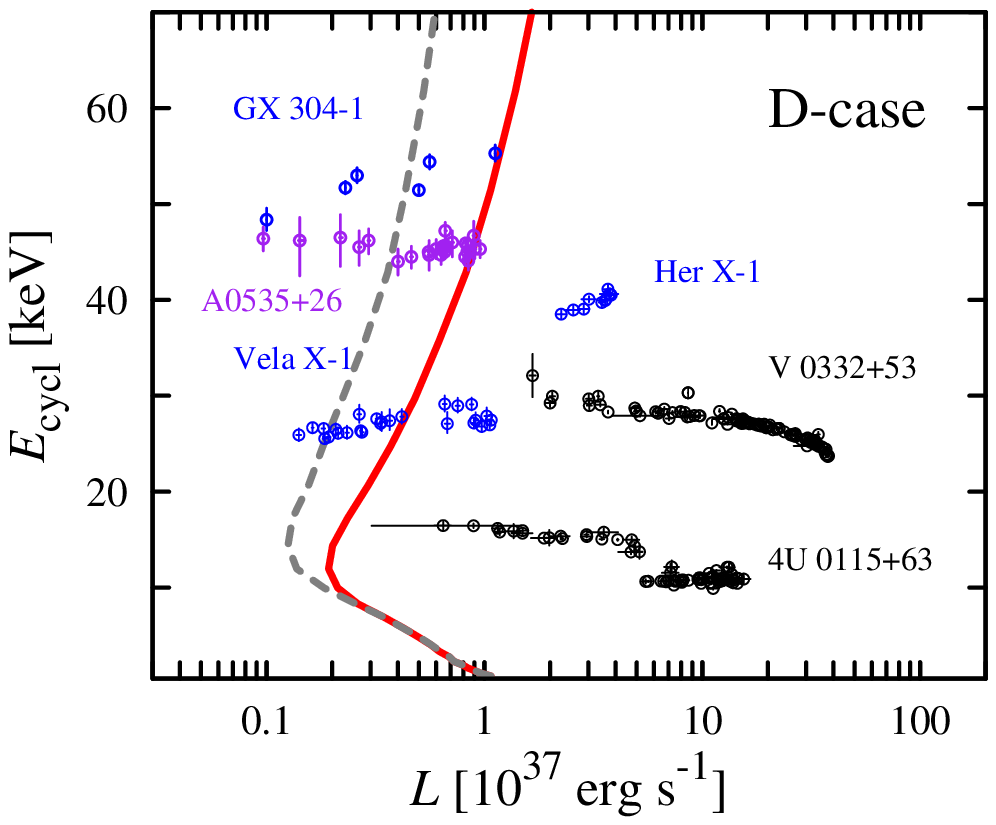} 
\caption{Comparison of the theoretical critical luminosity--$E_{\rm cycl}$ dependence to the data. 
In the W-case (left): matter fills the whole funnel cavity, while in the D-case (right) matter is confined a narrow wall of magnetic funnel. 
We use $\Lambda=0.5$ and $l_0/l=0.5$ to define the accretion channel geometry in the D-case. 
The grey and red curves show the theoretical dependences for pure O- and X-mode polarizations, respectively.
The data are plotted by blue circles for pulsars which show subcritical luminosity behaviour and 
by black circles for pulsars which supercritical behaviour. 
For Vela~X-1, the energy of the first harmonic divided by two is used. 
Other parameters are $M=1.4 {\rm M}_{\odot}$, $R=10^{6}$~cm,  $T_{\rm e}=1$~keV. 
}
\label{pic:WD_cases} 
\end{figure*}

Let us first demonstrate why it is important to compute Compton scattering 
cross-section accurately in order to get the correct behaviour of the critical luminosity 
on the $B$-field strength. For Thomson cross-section, 
the dependence is monotonic  (see dotted black curve in Fig.~\ref{pic:param}a) 
and just reflects the fact that at higher $B$ the spot area becomes smaller. 
The effect of the resonances can be demonstrated if we ignore them and take 
the cross section in the following form \citep{BS1975}: 
\beq \label{cross_wr} 
\begin{array}{ll}
\sigma_{\rm X} = \strut\displaystyle  \sigma_{\rm T} \left(\frac{E}{E_{\rm cycl}}\right)^2 , & E<E_{\rm cycl}, \\ 
\sigma_{\rm O} = \displaystyle \sigma_{\rm T} \left(\sin ^2\theta+\left(\frac{E}{E_{\rm cycl}}\right)^2\right) ,  & E<E_{\rm cycl}, \\ 
\sigma_{\rm X} = \sigma_{\rm O}=\sigma_{\rm T} ,    &   E\geq E_{\rm cycl} . 
\end{array}
\eeq
The solid blue and the red dashed curves in Fig.~\ref{pic:param}(a) show the dependences 
of the critical luminosity on $B$ computed with resonances and without them. 
We see that the dependence without resonances is much smoother. 
At small as well as large $B$, the curves coincide, just because 
most of the photons here come outside of the resonances. However, as soon as 
$k T_{\rm eff}\approx E_{\rm cycl}$, resonances increase the 
effective cross-section and reduce thus the critical luminosity.

Accretion flow geometry is a source of principal uncertainties. 
It affects the shape of the hot spot, its effective temperature and thereby
defines the effective cross section. 
The critical luminosities for  the  W- and  D-cases are compared 
in  Fig.~\ref{pic:param}(b). We see that qualitatively the behaviour is very similar. 
The effect of the pulsar inclination reflected in different length of the annular arc $l_0$ 
is shown in Fig.~\ref{pic:param}(c). 
The temperature of the plasma influences the depth of the dip in the critical luminosity 
around $E_{\rm cycl}\sim10$~keV (see Fig.~\ref{pic:param}d) affecting the width the resonance. 
Similarly, the correction to the effective temperature $\zeta$ because of the high plasma velocity 
shifts the resonance position and changes slightly the shape of the curve (see Fig.~\ref{pic:param}f). 
The effect of polarization is very strong at high $B$ (see Fig.~\ref{pic:param}e) because 
of the very different behaviour of the cross-section of the X- and O-mode below the  first resonance (se equation~(\ref{cross_wr})).
At low $B$ the cross-sections above the resonance are similar and therefore the critical luminosities coincide. 
In reality the polarization composition could be a function of the photon energy and the 
optical depth of a given point \citep{Miller1995}.

Our main conclusion that comes from Fig.~\ref{pic:param} is that the behaviour of the critical luminosity   
on the $B$-field strength is very robust. It has a minimum of a few times $10^{36}\ergs$ for $E_{\rm cycl}\sim$10--20 keV and 
increases to $\sim10^{37}\ergs$ for low $B$ nearly independent of the parameters. 
For the X-mode, the critical luminosities  increases sharply 
towards large $B$ owning to the drop of the Compton scattering cross-section below the resonance,
while the increase is less dramatic for the O-mode.

\subsection{Comparison with the data}
\label{sec:ThAndObs}

For a given set of parameters we are able to calculate the critical
luminosity as a function of the $B$-field strength 
(or, alternatively, of the cyclotron energy). As was discussed in the Introduction, 
the obtained dependence should separate the sources with hotspots on
the NS surface from the sources with accretion columns. 
Therefore, it is expected that the sources with luminosities
to the left from the curve show a positive correlation between cyclotron
line centroid energy and luminosity while the sources with
luminosities to the right should show a negative correlation.

In order to verify our theoretical predictions we use the observations
of five X-ray pulsars for which the dependence of cyclotron energy 
(or first harmonic energy as it is for Vela~X-1) on
luminosity is firmly established: GX 304--1 \citep{Kloch2012}, Her X-1
(\citealt{Staub2007}; \citealt*{VK2011}), V 0332+53 \citep{TsLS2010}, 4U 0115+63
\citep{TsL2007}, A 0535+26 \citep{Cab2007} and Vela X-1 \citep{NuSTARvela2014}.  It is believed that 
in all sources except Vela X-1 the mass accretion occurs mainly through the disc.
Vela~X-1 belongs to systems where accretion process occurs through the wind.
It is also an exceptional case since a clear increase of the first harmonic energy
with luminosity is visible, while the evolution of the energy of fundamental line 
with the luminosity is difficult to interpret. Thus,  for Vela~X-1 
we use the energy of the harmonic divided by two on the plots.

Two sources -- V~0332+53 with a confident negative correlation \citep{TsLS2010} and
4U 0115+63 with probable negative correlation
(\citealt{TsL2007,Muller2013}; \citealt*{Boldin2013}) --  clearly belong to the area of supercritical accretion.  
One source, GX 304--1, which shows a positive
correlation \citep{Yam2011,Kloch2012}, belongs to the area of subcritical accretion. 
The recent \textit{NuSTAR} observations of Vela X-1  show some hints on the
positive correlation between the position of first harmonic of the
cyclotron line and luminosity \citep{NuSTARvela2014}. 
It is likely that the critical luminosity of Vela~X-1 is around $10^{37}\,\ergs$. 
A~0535+26 which does not show positive or negative correlation \citep{Cab2007}
also belongs to the area of subcritical accretion.

The theoretical critical luminosity versus observed cyclotron energy curves 
for the wind and the disc accretion cases are shown in Fig.~\ref{pic:WD_cases} together with the data. 
We see that models well describe the data separating the two regimes, subcritical and supercritical, 
where the correlation changes from positive to the negative one. 
The X-mode polarization model is clearly preferred.

Her X-1, which probably shows a positive correlation \citep{Staub2007,VK2011}, 
should belong to the region of subcritical accretion as well. It is slightly off our relation. 
The critical luminosity for Her X-1 seems a bit higher than the 
predicted one, possibly because of the strong non-dipole $B$-field component, which
leads to a larger base area of the accretion channel  \citep*{ShPPr1991}. 
Alternatively, the data on Her X-1 can be explained if we assume $\Lambda=0.1$ as 
proposed by \citet{Becker2012}. This would shift the critical luminosity curve to the right 
by a factor of $\sim 2.5$. Such a small $\Lambda$, however, contradicts theories of 
disc--	magnetosphere interaction \citep[see e.g.][]{GL1979,Lai2014}.

\begin{figure}
\centering 
\includegraphics[width=8cm]{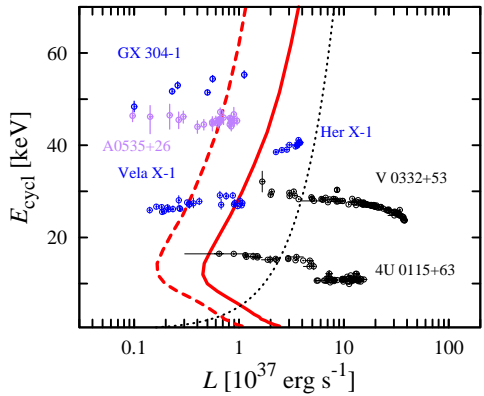} 
\caption{Red solid and dashed curves corresponds to the disc-accretion case
with different parameters. The dashed curves is for $l_0/l=0.5$ and
  mixed polarization  (the flux in X- and O-modes are equal), while the solid curve is for 
 $l_0/l=1.0$ and pure X-mode  polarization. In both cases $T_{\rm e}=1$ keV, $\Lambda=0.5$ and $\zeta=1.5$.
  The real critical luminosity value lies likely between the two  curves.  The predictions by 
  \citet{Becker2012} are shown with the black dotted line. 
  For the case of Vela~X-1 the first harmonic energy divided by two is presented. 
 Other parameters: $M=1.4 {\rm M}_{\odot}$, $R=10^{6}$~cm, $T_{\rm e}=1$~keV.
  }
\label{pic:final}
\end{figure}

Fig.~\ref{pic:final} demonstrates our theoretical curves for two possible polarization mixtures along with
the data. We see that the model for pure X-mode describes the data better. 
For comparison we also show the prediction of the model by \citet{Becker2012} (black dotted curve) 
which contradicts the behaviour of V~0332+53 and 4U 0115+63.

\section{Summary}
\label{discussion}

Following the theoretical model by \citet{BS1975, BS1976} we have
calculated the critical luminosity for the magnetized NS  as
a function of magnetic field strength (or, equivalently, as a function of the cyclotron energy).  
For the first time the exact effective cross section for Compton scattering 
in a strong magnetic field \citep{Harding1986} including the resonances was used. 
We have investigated the dependence of the results on the  
polarization composition,  the geometry of the accretion flow and the temperature of the accreting gas.  

We  showed that $L^*$ is not a monotonic function of the field
strength and reaches its minimal value of a few $\times 10^{36}\ergs$ 
around the observed cyclotron energy $\sim$10~keV that corresponds to the
surface magnetic field strength of $B\sim10^{12}$~G. 
Such critical luminosity is reached when a considerable amount of photons
from a hotspot have energy close to the cyclotron resonance and the
effective cross section reaches its maximum.  Since the typical hotspot
temperature is a few keV, the critical luminosity reaches its
minimum for the sources 
which have the cyclotron line close to the spectral peak. 
The critical luminosity increases for  small $B$ to  $L^*\sim 10^{37}\ergs$ nearly 
independent of the parameters. It also increases at large $B$-field strength
because photons have much smaller cross section below the cyclotron energy.
This behaviour is very robust and depends very little on the details of the model.

The obtained dependence of the critical luminosity on the $B$-field
strength should separate sources in subcritical regime of accretion
with a hotspot on the NS surface from supercritical regime with an
accretion column. Therefore, we expect to observe a positive and
a negative correlation between the cyclotron line centroid energy and
the luminosity for the sources below and above the critical luminosity, respectively.  

The comparison between the theoretical results and the data gives us an opportunity to obtain  
important parameters describing the accretion process onto a magnetised NS and provides 
additional method of diagnostics for systems with accreting NS.
The expected appearance in the near future of the high quality data
from the currently operating and planned X-ray telescopes will provide
an excellent opportunity to verify the proposed theoretical
predictions and to constrain some key parameters.
 
\section*{Acknowledgements}

This research was supported by the Magnus Ehrnrooth Foundation (VFS and JP), 
 the Jenny and Antti Wihuri Foundation (VFS and JP), 
the Russian Scientific Foundation grant 14-12-01287 (AAM), 
the Academy of Finland grant 268740 (JP), 
the German research Foundation (DFG) grant SFB/Transregio 7 "Gravitational Wave Astronomy" 
and the Russian Foundation of Fundamental Research grant 12-02-97006-r-povolzhe-a (VFS),
the grant RFBR 12-02-01265 (SST). We are grateful to Dmitrij Nagirner for a number of useful comments.


\label{lastpage}

\end{document}